\documentclass[aps,prb,twocolumn,showpacs,superscriptaddress,citeautoscript]{revtex4}
\usepackage{epsfig,graphicx}
\usepackage{times}

\newcommand{\kf}{k_F}

\newcommand{\br}{{\bf r}}

\newcommand{\bp}{{\bf p}}
\newcommand{\bq}{{\bf q}}

\newcommand{\al}{\alpha}
\newcommand{\be}{\beta}
\newcommand{\s}{\sigma}
\newcommand{\e}{\epsilon}

\newcommand{\Uext}{U_{\rm ext}}

\newcommand{\Utf}{U_{\rm TF}}

\newcommand{\Vrpa}{V_{\rm RPA}}
\newcommand{\hVrpa}{\hat V_{\rm RPA}}
\newcommand{\Vtf}{V^0_{\rm RPA}}
\newcommand{\hVtf}{\hat V^0_{\rm RPA}}

\newcommand{\Eop}{E_{\rm 1p}}
\newcommand{\Eri}{E_{\rm ri}}

\newcommand{\Fks}{{\cal F}_{\rm KS}}

\newcommand{\Ttf}{{\cal T}_{\rm TF}}
\newcommand{\Tks}{{\cal T}_{\rm KS}}

\newcommand{\Exc}{{\cal E}_{\rm xc}}
\newcommand{\Etot}{{\cal E}_{\rm tot}}

\newcommand{\nTF}{n_{\rm TF}}

\begin{document}

\title{Interactions and Broken Time-Reversal Symmetry in Chaotic Quantum Dots}

\author{Denis Ullmo}

\thanks{Permanent address: Laboratoire de Physique Th\'eorique et
  Mod\`eles Statistiques
(LPTMS), 91405 Orsay Cedex, France}

\affiliation{Department of Physics, Duke University, 
Durham, North Carolina  27708-0305}

\author{Hong Jiang} 

\thanks{Current address: Institut f\"ur Theoretische Physik,
J.W.Goethe-Universit\"at,Frankfurt am Main, Germany}

\affiliation{Department of Physics, Duke University, Durham, North
Carolina 27708-0305}

\affiliation{Department of Chemistry, Duke University, Durham, 
North Carolina 27708-0354}

\author{Weitao Yang}

\affiliation{Department of Chemistry, Duke University, Durham, 
North Carolina 27708-0354}

\author{Harold U. Baranger}
\affiliation{Department of Physics, Duke University, 
Durham, North Carolina  27708-0305}

\date{\today}

\begin{abstract}
When treating interactions in quantum dots within a
RPA-like approach, time-reversal symmetry plays an important role as
higher-order terms -- the Cooper series -- need to be included when this
symmetry is present.  Here we consider model quantum dots in a
magnetic field weak enough to leave the dynamics of the dot chaotic,
but strong enough to break time-reversal symmetry. The ground state
spin and addition energy for dots containing 120 to 200 electrons are
found using local spin density functional theory, and we compare the
corresponding distributions with those derived from an RPA-like
treatment of the interactions.  The agreement between the two
approaches is very good, significantly better than for analogous
calculations in the presence of time-reversal symmetry.  This
demonstrates that the discrepancies between the two approaches in the
time-reversal symmetric case indeed originate from the Cooper channel,
indicating that these higher-order terms might not be properly taken
into account in the spin density functional calculations.
\end{abstract}



\maketitle

The impressive progress seen in the recent years in the fabrication
and manipulation of semiconductor quantum dots \cite{Kouwenhoven97}
has been associated with a shift in the perspective with which these
nanoscopic objects are considered.  While the focus used to be on the
study of their basic properties, most recent experiments involve them
as part of a more complex system in which they play a specific role.
Some examples include spin filtering \cite{PotokMarcus03}, charge
detection \cite{Buks98}, or the manipulation of quantum
information \cite{Hayashi04}.

This new stage of maturity of quantum dot physics makes it necessary
to go beyond a theoretical understanding of the qualitative properties
of these systems, and to also develop good simulation tools which
allow one to predict their behavior quantitatively.  A natural way to
proceed is to rely on density functional theory
\cite{ParrYangBook,Jones89RMP}.  More precisely, since the spin turns
out to be an essential degree of freedom, we have in mind a spin
density functional theory, in the local spin density approximation
(LSDA), where each spin density is taken as an independent variable.
The properties of relatively small dots have been investigated in
great detail \cite{Reimann02RMP} in this framework, and more recently
this approach has been used for significantly larger dots containing
several hundreds of electrons \cite{Jiang03prl,Jiang04prb}.

In this large dot limit, and assuming chaotic dynamics within the
dots, it is expected that the fluctuations of experimentally relevant
quantities, such as the addition energy or the ground state spin,
should be reliably predicted within a Landau Fermi-liquid {\em
picture} of the Coulomb interaction and a model of wave functions and
eigenenergies based on Random Matrix Theory (RMT) and Random Plane
Waves (RPW) \cite{Blanter97,Ullmo01prbb,Usaj01,Usaj02,Aleiner02}.  (By
``fluctuations'' we mean the dot-to-dot variation or variation as a
function of energy typically seen in mesoscopic
systems.\cite{Kouwenhoven97}) We shall below refer to these
predictions (somewhat inappropriately) as the ``RPA approach'' since
the Landau Fermi-liquid picture arises from a Random Phase
Approximation to the screened Coulomb interaction
\cite{Blanter97,Aleiner02}. Quite unexpectedly, however, for a model
chaotic quantum dot the same quantities computed with LSDA turned out
to differ from these theoretical predictions, even at the qualitative
level \cite{Jiang03prl,Jiang04prb}.

This motivated us to introduce a second-order Strutinsky approximation
to the LSDA calculations;\cite{Ullmo04prb} it was shown to be precise
enough to serve as a basis for interpreting the LSDA results.  This
analysis indicated that the origin of the discrepancies between the
LSDA and RPA results arises from two reasons (on which we shall
elaborate  below): (i) an effective residual interaction which is
slightly larger in LSDA; (ii) the absence in LSDA of renormalization
of the Cooper channel associated with higher-order terms in the
residual interaction. By the Cooper channel we mean the interaction of two electrons following time-reversed paths for which, then, multiple scattering is highly likely \cite{note0}. 

Because the Cooper channel contribution is non-zero only for {\em
time-reversal invariant} systems, a direct implication of the above
analysis is that the agreement between LSDA simulation and RPA
prediction should be {\em significantly better} in the presence of a
magnetic field strong enough to break time-reversal symmetry.  By
showing that this prediction actually holds, the
goal of this paper is to provide an unambiguous demonstration of
the importance of the Cooper channel, and as a consequence to indicate
a possible limitation of the density functional approach to low
temperature quantum dot properties.

We shall therefore compare in this paper the distribution of addition
energies and ground state spins obtained in two different ways.  The
first one corresponds to a full fledged LSDA calculation for a
specific model quantum dot. The second is a prediction for a {\em
generic} chaotic system (i.e.\ not associated to any specific
confinement) for which an RPA-like treatment of the interaction is
used.

The model quantum dot for the LSDA calculations is
confined by the two dimensional potential  [${\bf r } = (x,y)$]
\begin{equation} \label{eq:qos}
  \Uext({\bf r}) = a \left[ \frac{x^4}{b} + b y^4 - 2\lambda x^2 y^2
    + \gamma(x^2y -y^2x)|{\bf r}| \right]
\end{equation}
and subjected to the uniform magnetic field $B \hat {\bf z}$. (We use
effective atomic units throughout the paper, which, for GaAs quantum
dots with effective electron mass $m^*=0.067 m_e$ and dielectric
constant $\epsilon=12.4$, corresponds to 10.08 meV for energy, and
10.95 nm for length.)  Here we use $a = 10^{-4}$, which for particle
number in the range $[120,200]$ gives a parameter $r_s \simeq 1.4$
(the ratio between the interaction energy and kinetic energy, formally
defined as $r_s=1/\sqrt{\pi n} a_0$ for 2D bulk systems).  The
coupling $\lambda $ between the two oscillators is fixed around $0.6$
so that the classical motion within the dot is in the hard chaos
regime.  Finally, the parameters $b$ and $\gamma$ eliminate any
discrete symmetry. We shall keep $b =\pi/4$ constant, but, to enhance
the statistical significance of the various properties studied, we
will use the five sets $(\lambda,\gamma) = \{ (0.53, 0.2),\, (0.565,
0.2),\, (0.6, 0.1),\, (0.635, 0.15) , \,(0.67, 0.1) \}$.

In the spin density functional description, one considers a functional
of both spin densities $[n^\alpha(\br),n^\beta(\br)]$
\begin{equation} \label{eq:Fks}
    \Fks[n^\al,n^\be]=\Tks [n^\al,n^\be]+\Etot[n^\al,n^\be] \; ,
\end{equation}
where $\s = \al,\be$ correspond to majority and minority spins,
respectively. The second term is an explicit
functional of the densities,
\begin{eqnarray} \label{eq:def_Etot}
 \Etot[n^\al,n^\be] \equiv \int \!d\br \;n(\br) \;\Uext(\br) \qquad
\qquad \qquad \nonumber\\ + \frac{e^2}{2}  \int \! d\br d\br' \,
\frac{n(\br) n(\br')}{|\br-\br'|}+\Exc [n^\al,n^\be] \; ,
\end{eqnarray}
where $n(\br)=n^\al(\br) + n^\be(\br)$, $\Uext(\br)$ is the exterior
confining potential Eq.~(\ref{eq:qos}), and we use the Tanatar-Ceperley
parametrization of the  the exchange-correlation
term $\Exc [n^\al,n^\be]$ \cite{Tanatar89,note2}. 

The kinetic energy term $\Tks [n^\al,n^\be]$, on the other hand, is
expressed in terms of a  set of auxiliary orthonormal functions
$\psi_i^{(\al,\be)}$ ($i=1,\cdots,N^{\al,\be}$)
as
\begin{equation}
    \Tks[n^\al,n^\be] 
            =  \!\sum_{\scriptstyle {\s=\al,\be} \atop \scriptstyle
            {i=1, N_\s}} 
\! \int \psi_i^{\s*}({\bf r}) {\left[ \bp - (e/c)A(\br)\right]^2\over
            2m} 
\psi_i^\s({\bf r}) d{\bf r},
    \label{eq:def_TKS}
\end{equation}
with ${\bf A}(\br)$ the vector potential generating the magnetic field
and $ n^\s({\bf r}) \equiv \sum_{i=1}^{N_\s} |\psi^\s_i({\bf r})|^2$.

The LSDA energy for a given value of the total number of particles
$N=N^\al + N^\be$ and spin $S=(N^\al - N^\be)/2$ is obtained by
minimizing the functional Eq.~(\ref{eq:Fks}) under the constraints
$\int d\br n^{\al,\be}(\br) = N^{\al,\be}$.  The ground state energy for
a given $N$ is then obtained by picking the spin $S$ with lowest
energy.  The details of the practical implementation of these
calculations can be found in Ref.\ \ \onlinecite{Jiang03prb}.

\begin{figure}
\includegraphics[width=2.5in,clip]{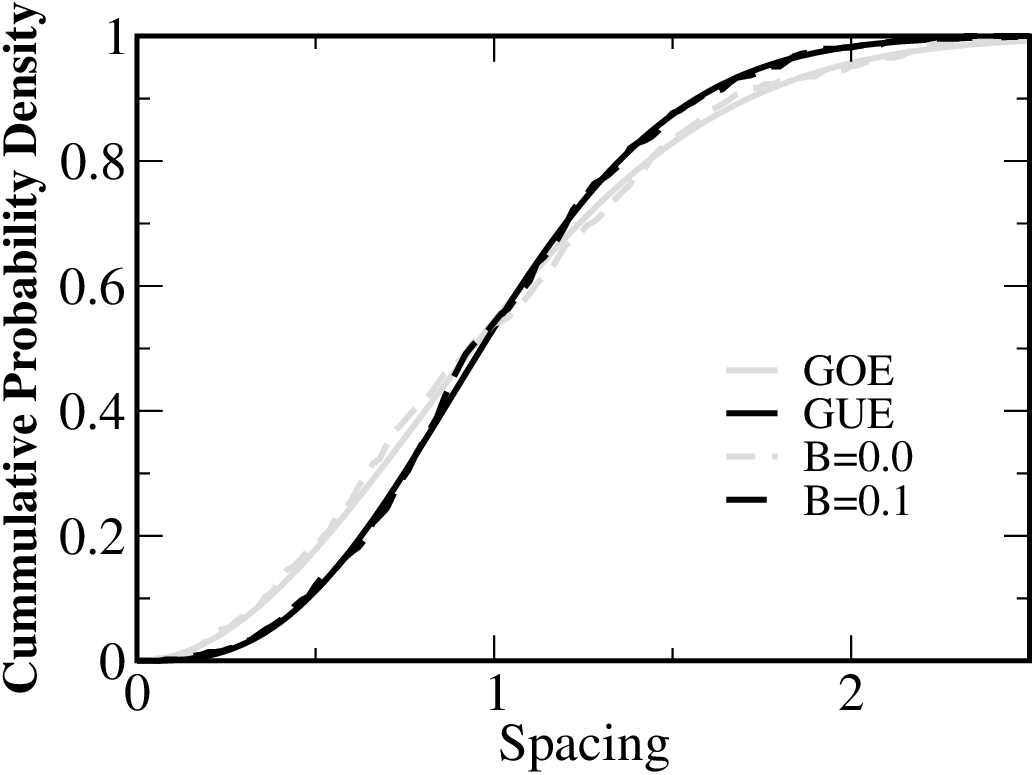}
\caption{Cumulative density of nearest neighbor level spacing for the
  effective Hamiltonian derived from the Thomas Fermi approximation at
  $B=0 {\rm T}$ (dashed gray) and $B=0.15 {\rm T}$ (dashed black).
  The solid lines are the corresponding Wigner surmise distributions
  for the orthogonal (gray) and unitary (dark) Gaussian ensembles. }
\label{fig:nns_Mij}
\end{figure}

Already quite good approximations of the
densities $\nTF^{\al,\be}(\br)$ are obtained via the Thomas-Fermi
approximation (see e.g.\ Fig.~2 in Ref.\ \ \onlinecite{Ullmo04prb}),
which amounts to replacing the Kohn-Sham kinetic energy
Eq.~(\ref{eq:def_TKS}) by the explicit density functional
\begin{equation} \label{eq:TTF} 
\Ttf  [n^\al,n^\be]  =  \frac{1}{2 N(0)} \int \! d \br
    \big[n^\al(\br)^2 + n^\be(\br)^2\big] \; ,
\end{equation}
(valid in 2D) with $N(0) \!=\! m/\pi \hbar^2$ the
two-dimensional density of states. Once the Thomas-Fermi calculation
has been performed (see Ref.\ \ \onlinecite{Jiang04jcp} for the practical
implementation), the (spin dependent) one-particle effective
Hamiltonian
\begin{equation} \label{eq:HTF}
  H^{\al,\be}_{\rm TF} = \left[ \bp - (e/c)A(\br) \right]^2/2m +
\Utf^{\al,\be}(\br)
\end{equation}
 defined in terms of the Thomas-Fermi self consistent potential
\begin{equation} \label{eq:Utf}
  \Utf^{\al,\be}(\br) =\frac{\delta \Etot}{\delta
  n^{\al,\be}(\br)}[\nTF^\al,\nTF^\be]
\end{equation}
gives a good account of the qualitative behavior of the dynamics
within the dot, including self-consistent effects associated with the
Coulomb interactions. We can therefore use $H_{\rm TF}$ to assess the
degree of chaoticity of the dots as well as the effectiveness of the
time-reversal breaking term.  Fig.~\ref{fig:nns_Mij} displays the
cumulative density of the nearest-neighbor level spacing for the
effective Thomas-Fermi Hamiltonian and for two values of the magnetic
field $B \!=\! 0$ and $B\!=\!0.15$~T. (The latter corresponds to about
7 flux quanta through the area of the dot, or a filling factor of
about 50 at $r_s \sim1.4$.)  The perfect agreement with the random
matrix theory prediction for the GOE and GUE ensembles, respectively,
shows both that the systems we consider are indeed chaotic and that
$B\!=\!0.15$~T is enough to break time-reversal invariance.  We have
performed additional checks on the statistics of the wavefunctions,
and in particular on the distribution of the quantities $M_{ij}$ and
$N_{ij}$ introduced in Eq.~(\ref{elabet}), which further confirms
this point.

\begin{figure}
\includegraphics[width=1.9in,clip]{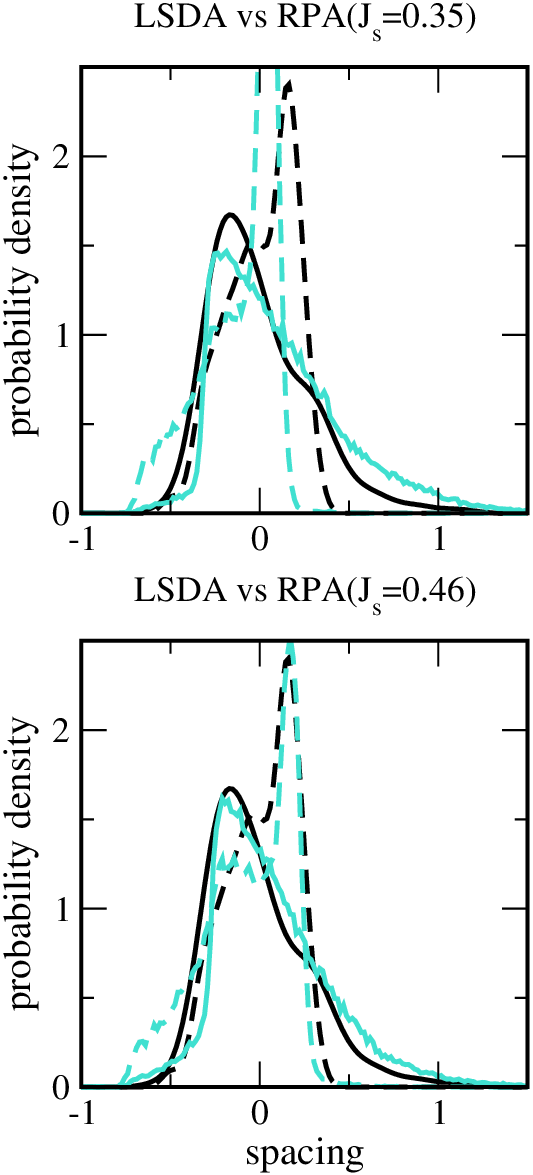}
\caption{(Color online) Addition spectra obtained from the  LSDA
  calculations (dark), compared with theoretical predictions derived
  from  RPA/RPW (lighter, blue online).  Solid:
  $N$ even; Dashed: $N$ odd.  Top: strength of the RPA
  interaction obtained by matching the QMC value, $J_s = f^a_0 = 0.35$ (for $r_s=1.4$).
  Bottom: better agreement is obtained with a slightly
  larger value of this parameter, $J_s = 0.46$.
  }
\label{fig:results:psp}
\end{figure}

\begin{figure*}[t]
\includegraphics[width=4.5in,clip]{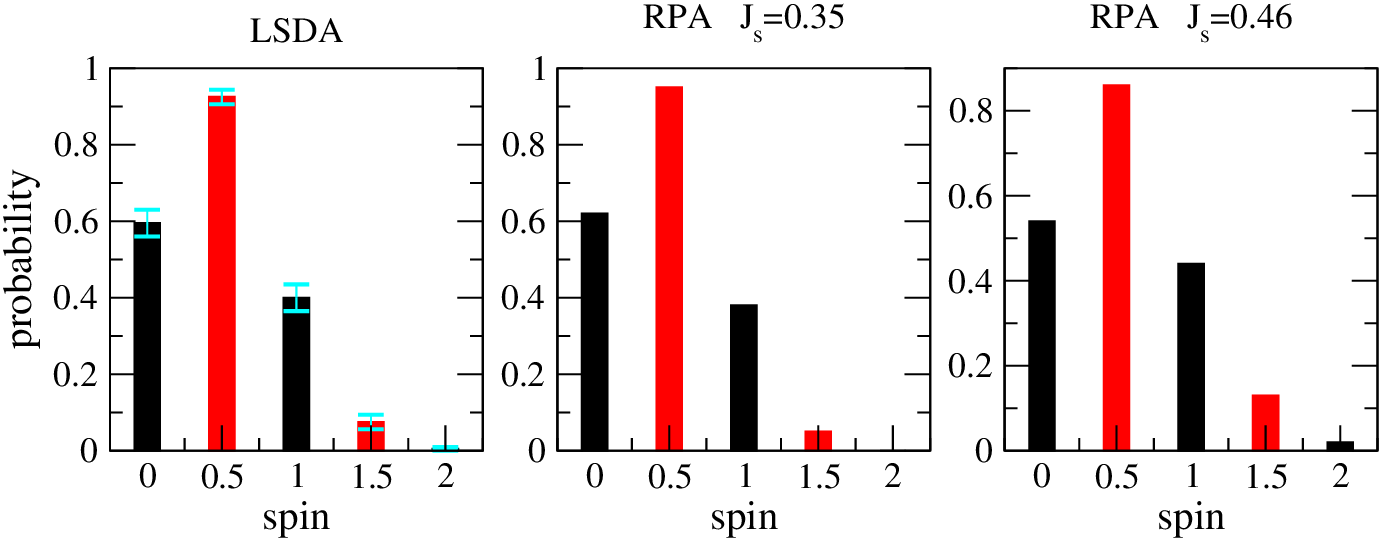}
\caption{(Color online) Spin distribution obtained from the LSDA
  calculations (left), the RPA/RPW model with $J_s = f^a_0 = 0.35$
  (middle), and $J_s = 0.46$ (right).  Contrary to the addition
  spectra, slightly enhancing the RPA interaction  does not
  improve the agreement with the LSDA simulation for the spin
  distribution; rather, it makes it worse.  }\label{fig:results:spin}
\end{figure*}

Being confident that the choice of parameters and magnetic field
corresponds to the regime we would like to study, we can now compute
the ground state energy and spin for our model quantum dot with the
full fledged LSDA self-consistent approach.  The
resulting distributions of addition energy and ground state spins,
for particle number $N$ in the range $[120,200]$, are shown in
Figs.~\ref{fig:results:psp} and \ref{fig:results:spin}.

We would like now to compare these distributions with theoretical
predictions in which (i) the interactions are treated by a RPA-like
approach and (ii) the statistical properties of the eigenlevels and
eigenfunctions are modeled in terms of RMT and RPW.
The RPA treatment of the interactions \cite{Blanter97,Aleiner02} is
essentially a Landau Fermi-Liquid picture in which the electrons are 
considered to be quasi-particles evolving under an effective 
one-particle Hamiltonian $H_{\rm eff}$ (which we can think of as being
$H_{\rm TF}$) and interacting through a weak short-range residual
interaction $\Vrpa(\br-\br')$. For {\em broken time-reversal
symmetry},  $\Vrpa(\br-\br')$ can be taken into account by first-order
perturbation theory.  Up to a smooth component in which we are not
interested, the energy of the quantum dot for a given set of
occupation numbers $f_{i}^{\s}=0,1$ of the orbitals $\psi_i$ of 
$H_{\rm eff}$ is therefore the sum of the one-particle energies
\begin{equation}
   \Eop = \sum f_{i}^{\s} \e_i \; 
\end{equation}
and the residual interaction direct plus exchange terms
\begin{equation} \label{elabet}
 \Eri  =  \frac{1}{2} \sum_{i,j,\s,\s'} f_{i}^{\s} f_{j}^{\s'}  M_{ij}
   -  \frac{1}{2} \sum_{i,j,\sigma}f_{i}^{\s} f_{j}^{\s} N_{ij}
\end{equation}        
with 
\begin{eqnarray*}
M_{ij}  &=& 
    \int d\br d\br' \left| \psi_i(\br) \right|^2
            \Vrpa(\br-\br') \left| \psi_{j}(\br') \right|^2  \\
 N_{ij}  &=&   
          \int d\br d\br' \psi_i(\br) \psi_{j}^*(\br)
            \Vrpa(\br-\br') \psi_{j}(\br') \psi_{i}^*(\br') \; .
\end{eqnarray*}

Note that once the interaction potential $\Vrpa(\br-\br')$ is
determined, all the relevant quantities are expressed in terms of the
eigenvalues $\e_i$ and eigenfunctions $\psi_i$ of $H_{\rm eff}$.  The
RMT/RPW model consists in fixing the eigenvalue fluctuations according
to RMT and those of the eigenfunctions by describing them as a
superposition of random plane waves, for levels within the Thouless
energy of the Fermi energy. This fixes the fluctuations of $ \Eop$ and
of the $M_{ij}$ and $N_{ij}$.

The weak interaction obtained from the RPA treatment
\cite{Blanter97,Aleiner02} is quite naturally the zero frequency and
low momentum approximation of the screened RPA interaction, which for
2D systems takes the form
\begin{eqnarray}
  \Vtf(\br) & = & \int \frac{d\bq}{(2\pi)^2} \ \hVtf (\bq) \
      \exp[i\bq\!\cdot\!\br] \ , \nonumber \\
  \hVtf (\bq) & = &
    \frac{N(0)^{-1}}{1+ r_s^{-1} (|\bq| /\kf)/\sqrt{2} } \ .
        \label{eq:Vtf}
\end{eqnarray}
It is known from quantum Monte Carlo (QMC) results that, in the regime
of interest, RPA slightly underestimates the interaction strength. Thus we use
a boosted interaction
  \begin{equation}
   \Vrpa = \xi \Vtf(\br)
  \end{equation}
with $\xi$ adjusted so that RPA matches QMC. Note that this is
completely consistent since the Tanatar-Ceperley parametrization of
LSDA comes from fitting the QMC results as well. The natural quantity
to match is the Fermi circle average
 \begin{equation}
	J_S \equiv \langle \hVrpa \rangle_{\rm fs} = 
	\!\int_0^{2\pi} \!\frac{ d\theta}{2\pi} \,
           \hVrpa \bigl(\kf \sqrt{2 (1+\cos\theta)} \bigr)
  \end{equation}
which should be interpreted as the Fermi-Liquid parameter $f^0_a$. For
the value $r_s=1.4$ considered here, QMC gives \cite{note1} $f^0_a =
0.35$ while $\langle \hVtf \rangle_{\rm fs} = 0.31$. Thus, for our
density, the natural value is $\xi=1.1$; we have used other values as
well in order to explore trends.


The distributions of addition energy and ground state spin for $\xi \!=\! 1.1$
and $1.5$, corresponding to $J_s  \!=\!  0.35$ ($ = f_a^0$) and $0.46$, are shown
in Figs.~\ref{fig:results:psp} and \ref{fig:results:spin}.  We observe
that, although the lower value of the interaction gives already
reasonable agreement for the addition energy distribution, a better
result is obtained, especially for $N$ odd, when the interaction is
enhanced.  In contrast, the spin distribution is accurately reproduced
by just taking $J_s = f_0^a$, while the stronger interaction predicts
slightly too many high spin ground states.

Despite the apparent contrast here, these results are nicely in line
with the analysis using the Strutinsky approximation in Ref.\ \
\onlinecite{Ullmo04prb}.  Indeed, the results there indicate that the
main source of discrepancy between LSDA and RPA/RPW comes from the
absence of screening of the Cooper channel in the former.  \textit{Since the
Cooper channel does not contribute when time-reversal symmetry is
broken, the much better agreement between LSDA and RPA/RPW here, as
compared to the time-reversal invariant situation considered before
\cite{Jiang03prl,Ullmo04prb}, demonstrates the importance of this higher-order interaction effect in the ground state properties of quantum dots.}  

It was also pointed out in Ref.\ \ \onlinecite{Ullmo04prb} that beyond the effect
of the Cooper channel, LSDA differed from RPA/RPW in two aspects:
(i)~a slightly larger effective residual interaction, and (ii)~the
existence of spin contamination, i.e.~of solutions of the Kohn-Sham
equations with different wave functions for the majority and minority
spins. The first of these points explains why it is necessary to
slightly enhance the RPA interaction to perfectly reproduce the LSDA
addition energy distribution.  The spin contamination mechanism, however,
has a tendency to produce slightly smaller ground state spins, and
therefore offsets the first point when the statistics of spin
ground states are considered.  As a consequence, spin statistics from
the LSDA calculations nicely agree with the RPA/RPW prediction without
enhancing the RPA interaction.

To conclude, we have shown that because of the absence of the Cooper
channel contribution when time-reversal symmetry is broken, the
addition energy and ground state spin statistics obtained from LSDA
calculations for a chaotic quartic oscillator at weak magnetic field
closely follow RPA/RPW predictions.  This demonstrates in an
unambiguous way that higher-order interaction effects in the Cooper channel
are indeed the main cause of the discrepancies observed between LSDA
and RPA/RPW in the time-reversal symmetric case.  The
remaining differences can be understood as arising from a combination
of spin contamination and the slightly larger effective interaction
that can be derived from LSDA within a second-order Strutinsky scheme.
Finally, we stress that beyond these differences, both methods predict
a great sensitivity of the addition energy distribution (at weak
magnetic field) to the precise value of the exchange parameter $J_s$.
We suggest that experimentally measuring these distributions, at a
temperature low enough\cite{Usaj01} to resolve the shift in the maxima between the
odd and even distributions, would provide a sensitive way to directly
access this quantity.

This work was supported in part by NSF Grant No. DMR-0103003.


\begin{thebibliography}{21}
\bibitem[{not({\natexlab{c}})}]{note1}
\bibinfo{note}{More precisely, this value corresponds to the parametrization of
  the exchange-correlation functional we have used in the LSDA calculations.}
\end{thebibliography}

\end{document}